\documentclass[12pt]{article}
\usepackage{epsfig,graphicx}

\begin{document}
\title{A model for the dynamics of extensible semiflexible polymers}
\author{
G.T. Barkema
\thanks{Institute for Theoretical Physics, Utrecht University,
P.O. Box 80195, 3508 TD  Utrecht, The Netherlands, and
Instituut-Lorentz for Theoretical Physics, Leiden University,
P.O. Box 9504, 2300 RA Leiden, The Netherlands},
and
J.M.J. van Leeuwen
\thanks{Instituut-Lorentz for Theoretical Physics, Leiden University,
P.O. Box 9504, 2300 RA Leiden, The Netherlands}}

\date{\today}
\maketitle

\begin{abstract}
We present a model for semiflexible polymers in Hamiltonian formulation
which interpolates between a Rouse chain and worm-like chain. Both
models are realized as limits for the parameters. The model parameters
can also be chosen to match the experimental force-extension curve
for double-stranded DNA. Near the ground state of the Hamiltonian,
the eigenvalues for the longitudinal (stretching) and the transversal
(bending) modes of a chain with $N$ springs, indexed by $p$, scale
as $\lambda^l_p\sim(p/N)^2$ and $\lambda^t_p\sim p^2(p-1)^2/N^4$
respectively for small $p$. We also show that the associated decay
times $\tau_p \sim (N/p)^4$ will not be observed if they exceed the
orientational time scale $\tau_r\sim N^3$ for an equally-long rigid rod,
as the driven decay is then washed out by diffusive motion.
\end{abstract}

%\pacs{36.20.-r,64.70.km,82.35.Lr}

\section{Introduction}

There are many models for the equilibrium behavior of long polymers but few for 
their dynamics. The celebrated Rouse model~\cite{rouse} describes the temporal 
behavior of the
chain by a set of (Rouse) modes in a harmonic potential. Such a Rouse (or fantom) 
chain is however too flexible to describe the behavior of biological polymers. For instance 
the persistence length of a Rouse chain is zero, while e.g. double-stranded (ds) DNA 
persists in its 
orientation over some hundred base pairs. It is easy to dress the Rouse model with a 
term that suppresses bending of the chain, while preserving the dynamics in the form
of a set of independent Rouse modes. However, the behavior of such a chain under 
an external force, i.e. the force extension curve, does still not correspond to that of the
semiflexible biopolymers, as it responds to the force as a harmonic spring, while 
the semiflexible chain has a characteristic rapid increase of the extension
for small forces, crossing over to a slow linear increase with the force. Clearly 
Rouse-type chains have  too little rigidity against longitudinal forces.

An opposite approach is to represent the biopolymer as a worm-like chain \cite{kratky,wilhelm}, 
which is a continuum description with the appropriate stiffness against bending. 
The worm-like chain (WLC) keeps a fixed contour length and only undergoes transversal 
undulations. A microscopic realization of the worm-like chain is the model in which the 
monomers are connected by rigid bonds. The bending stiffness is incorporated in the 
hamiltonian by a term suppressing the angle between two successive bonds. 
The rigidity of the bonds guarantees the invariance of the contour-length under the 
possible motion of the monomers. We refer to this model as the ``jointed chain'' 
(in analogy with the well known freely-jointed chain \cite{freely}). A refinement is to let the 
bond length vary with the external force. Such an ``extensible jointed chain'' can very well 
reproduce the measured force-extension curve of bio-polymers by an appropriate choice 
of the bending stiffness. A model close to the extensible jointed chain has recently been
introduced by Kierfeld et al. \cite{lipowsky},  who extensively discuss its equilibrium properties. 

However, the dynamics of jointed chains is quite involved, as the motion of the
monomers is strongly constrained by the rigidity of the bonds, which introduces 
unphysical long-range correlation. E.g. the motion between the end monomers 
of the chain is correlated due to the constant length of the contour. 

Thus a polymer model that can treat the dynamics of biological polymer fragments, 
such as occuring in networks, is quite welcome. These fragments are of the order 
of the persistence length, rendering a continuum description not adequate and a 
description on base-pair level necessary. 
In this paper we propose such a model combining  to a large extend the mathematical 
simplicity of  the Rouse model, while describing accurately the experimental 
force-extension curve. 

The hamiltonian for our flexible and extensible chain reads:
\begin{equation} \label{a1}
{\cal H} ={\lambda \over 2} \sum^N_{n=1} (|{\bf u}_n|-d)^2 - 
 \kappa \sum^{N-1}_{n=1} {\bf u}_n \cdot {\bf u}_{n+1} -{\bf F} \cdot {\bf L}.
\end{equation}  
Here ${\bf u}_n$ is the bond vector between monomer $n-1$ and $n$ and ${\bf L}$ the
end-to-end vector
\begin{equation} \label{a2}
{\bf u}_n = {\bf r}_n - {\bf r}_{n-1}, \quad \quad \quad 
{\bf L} = {\bf r}_N - {\bf r}_0 =  \sum^N_{n=1} {\bf u}_n.
\end{equation} 
${\bf r}_n$ is the position of the $n$-th monomer ($n=0,1, \cdots , N$).  ${\bf F}$ is a force
tending to orient and stretch the chain. The first term in (\ref{a1}) is a single-bond
harmonic interaction, providing the length scale  $d$ and the energy parameter 
$\lambda$. The second term is a nearest-neighbor-bond interaction giving the model a 
bending stiffness measured by $\kappa$. The above defined hamiltonian is the simplest 
of a class of hamiltonians with bond interactions, not only between nearest neighbors 
as in (\ref{a1}), but also longer-ranged interactions. 
If we set $d=0$ the model reduces to that of  Marques and Frederickson \cite{marques} 
(with slightly different boundary conditions).  As we mentioned above, such a Rouse-like 
model does not have an acceptable force-extension curve. We will argue that a  finite 
$d$ changes the  properties from being rather unrealistic to realistic. The difference 
between our model and that of Kierfeld et al.\cite{lipowsky} is minor as far as the 
equilibrium properties are concerned. However the difference is major with respect to 
the dynamics. While in \cite{lipowsky} the bending energy only depends on the angle 
between the bonds, in our case it depends on the bond vectors. The angles are
dynamically very inconvenient parameters: changing one angle while keeping the
others fixed, involves the motion of a whole segment of the chain.
In our model monomer positions can be changed independently, only changing the
incident bond vectors of that monomer. 

In this paper we discuss the mechanical and statistical properties of the model. 
We formulate the dynamical equations for the monomer positions, as well
as those for an equivalent description in terms of the modes of the system.
We also give a few implications of the mode equations.
In a subsequent study we design an efficient scheme for the solution of the
dynamical equations and report on extensive simulations of the model.

\section{The parameters of the model}

By varying the parameters $d, \lambda$ and $\kappa$ we cover a wide range of 
physical systems. We can eliminate the value of  a finite $d$ from the problem by  scaling 
${\bf u}_n$ with $d$. Then the hamiltonian reads
\begin{equation} \label{a3}
{\cal H} = {\lambda d^2 \over 2} \sum^N_{n=1} (|{\bf u}_n|-1)^2 - 
\kappa d^2 \sum^{N-1}_{n=1} {\bf u}_n \cdot {\bf u}_{n+1} - 
d {\bf F} \cdot \sum^N_{n=1} {\bf u}_n.
\end{equation}
Thus $d$ can be absorbed in the parameters $\lambda, \kappa$ and $F$. Therefore $d$
does not show up directly in the expressions for the static and dynamic properties. 
To translate the properties to a real chain one has to multiply the bond vectors with a 
well chosen $d$. The form (\ref{a3}) shows that $d=0$ is a singular point of the 
hamiltonian (\ref{a1}). Amongst others, the case $d=0$
has a lowest energy state in which all bond lengths vanish. 

As our main interest is in describing biological polymers, which have  a fairly rigid bond length 
and a large persistence length, we consider large values of $\lambda$ and $\kappa$.
So it is useful to consider the ratio between $\lambda$ and $\kappa$ as characteristic 
for the model. As we will see their common magnitude couples to the temperature. 
So we write 
\begin{equation} \label{a4}
\kappa = \nu \lambda
\end{equation} 
and the hamiltonian gets the form that we will use in this paper:
\begin{equation} \label{a5}
{\cal H}^* =\frac{\cal H}{\lambda d^2} = {1 \over 2} \left(\sum^N_{n=1}
 (|{\bf u}_n|-1)^2 - 2 \nu \sum^{N-1}_{n=1} {\bf u}_n \cdot {\bf u}_{n+1}\right) -
 {\bf f} \cdot  \sum^N_{n=1} {\bf u}_n,
\end{equation} 
with the abbreviation 
\begin{equation} \label{a6}
{\bf f} = {\bf F} /(\lambda d).
\end{equation}   
Equilibrium is governed by the ratio ${\cal H}/k_B T$, which we write as
\begin{equation} \label{a7}
{{\cal H} \over k_B T} = \frac{{\cal H}^*}{T^*} 
\end{equation} 
where the reduced temperature $T^*$ equals
\begin{equation} \label{a8}
T^* = k_B T/ (\lambda d^2).
\end{equation} 
Thus the combination $\lambda d^2$ is absorbed in the reduced temperature $T^*$.
We will discuss the model in terms of the dimensionless parameters $T^*, \nu$ and $f$.
Large values of $\lambda$ manifest themselves in low values of $T^*$, implying a
fairly constant bond length. Assuming a constant bond length reduces the model to 
the jointed chain and indeed the force-extension curve of our model closely follows that
of the extensible jointed chain. 

For the analysis of the model we split the hamiltonian in three parts according to the 
power in which the bond vectors (or monomer positions) occur:
\begin{equation} \label{a9}
{\cal H}^* = {\cal H}^*_h +{\cal H}^*_s + N /2.
\end{equation}
The first term, containing the quadratic terms in the bond vectors, has the form
\begin{equation} \label{a10}
{\cal H}^*_h= {1 \over 2}\sum^N_n u^2_n - \nu \sum^{N-1}_{n=1} {\bf u}_n \cdot {\bf u}_{n+1}. 
\end{equation} 
It is instructive to compare this hamiltonian with the one used by Marques and 
Frederickson \cite{marques}, which, using our parameters, reads
\begin{equation} \label{a11}
{\cal H}^*_h= {1-2\nu \over 2}\sum^N_n u^2_n + \frac{\nu}{2}
\sum^{N-1}_{n=1} ({\bf u}_n - {\bf u}_{n+1})^2. 
\end{equation} 
One observes that stability requires the bound $\nu < 1/2$, which we also have to
impose on (\ref{a5}). 
Close inspection shows that (\ref{a10}) and (\ref{a11}) are the same, apart from two 
boundary terms: $u^2_1$ and $u^2_N$ occur in (\ref{a10}) with the coefficient 
$1/2$, while in (\ref{a11}) they have the weight $(1-\nu)/2$. For the polymer properties
this difference is of little importance, but the form (\ref{a10}) permits exact 
diagonalization in closed form, while for (\ref{a11}) one has to rely on numerical 
diagonalization.  We will use ${\cal H}^*_h$ as expressed in terms of the monomer 
positions 
\begin{equation} \label{a12}
{\cal H}^*_h= {1 \over 2}\sum^N_{m,n} H_{m,n} \, {\bf r}_m \cdot {\bf r}_n.
\end{equation} 
We will show that the matrix $H_{m,n}$ can be exactly diagonalized. It would be 
the full (force-free) hamiltonian if we were to set $d=0$.

The term ${\cal H}_s$ represents the stabilizing force and has the form
\begin{equation} \label{a13}
{\cal H}^*_s = -\sum^N_{n=1}\,  [u_n +{\bf f} \cdot {\bf u}_n].
\end{equation} 
The first term in (\ref{a13}), involving the contour length of the chain, plays an 
essential role in stabilizing the chain. It prevents however an exact 
diagonalization of the model, which is possible for the harmonic terms. 
The second term in (\ref{a13}) accounts for the influence of the force on the chain. 

The last term in (\ref{a9}) is a trivial constant having no influence, neither on the 
equilibrium nor on the dynamics; we will pay no further attention to it. 

\section{Ground state Properties}

The dynamics of the model is our main interest. There are a number of  equilibrium
properties which are important for the dynamical analysis. We summarize them
here.

\subsection{ The ground state configuration} 

The ground state or lowest energy configuration prevails at low $T^*$.
With a non-vanishing external force, the chain starts to align itself with the force and the 
configuration becomes a straight line. All bond vectors point in the direction of ${\bf f}$
and may be written as ${\bf u}_n = u_n \hat{\bf f}$. The values of the length $u_n$ 
follow from minimization of the energy, leading to the equations
\begin{equation} \label{b1} 
\left( \begin{array}{rcl}
    u_1 - \nu u_2  & = & 1+f,\\*[2mm]
    -\nu u_1 + u_2 - \nu u_3 & = & 1+f, \\*[2mm]
     \cdots &  & \\*[2mm]
    -\nu u_{N-2} + u_{N-1} - \nu u_N & = & 1+f, \\*[2mm]
    - \nu u_{N-1} + u_N & = & 1+f.
\end{array} \right)
\end{equation} 
Note that the solution for $f \neq 0$ follows from  that for $f=0$ by multiplying all $u_n$
by $1+f$. A set of equations of the type (\ref{b1}) can be solved by the  standard technique
of the generating function. The solution reads
\begin{equation} \label{b2}
  u_n = \frac{1+f}{1-2\nu} \left[1-\frac{\cosh\{\alpha(N+1-2n)\}}{\cosh\{\alpha(N+1)\}} \right],
\end{equation}
 with $\cosh (2 \alpha)= 2 /\nu $. For the bulk value we take the bond $n=N/2$ in the middle.
The hyperbolic cosine in the numerator then obtains the argument $\alpha$, while the
denominator is exponentially large in $N$. Leaving out exponentially small contributions the
bulk values equals
\begin{equation} \label{b3}
b(f) =  \frac{1+f}{1-2\nu}.
\end{equation} 
The bonds become somewhat shorter near the ends of the chain. The value $b=b(0)$ is an 
important (numerical) value by which $d$ has to be multiplied in order to match the 
experimental bond length $a$. In other words $d=a/b$. We find that for small $T^*$, 
$b$ is still close to the value (\ref{b3}). 

\subsection{ Eigenmodes near the ground state}\label{modes}

The ground state energy responds harmonically to small deviations, yielding a
quadratic increase in the energy. We first discuss the force free case.
One can find the Hessian matrix by twice differentiating the Hamiltonian with
respect to the positions of the monomers. As the dependence on the monomer
 positions is exclusively through the bond vectors, we use the rule
\begin{equation} \label{b4}
\frac{\partial }{\partial {\bf r}_n} = \frac{\partial }{\partial {\bf u}_n} -
\frac{\partial }{\partial {\bf u}_{n+1}}, 
\end{equation} 
and define the bond matrix
\begin{equation} \label{b5}
{\bf B}_{m,n} =  {\partial^2 {\cal H}^* \over \partial {\bf u}_m  \partial {\bf u}_n}. 
\end{equation} 
The indices in the matrix $B$ run between $1 \leq m,n \leq N$. We extend the matrix by 
putting the elements equal to zero, whenever one of the indices equals 0 or $N+1$. 
The Hessian matrix $M$ is then obtained as 
\begin{equation} \label{b6}
{\bf M}_{m,n} = {\bf B}_{m,n}- {\bf B}_{m,(n+1)} - {\bf B}_{(m+1),n} + {\bf B}_{(m+1),(n+1)}.
\end{equation} 
$M_{m,n}$ runs in the interval $0 \leq m,n \leq N$. 

Let us first consider the part of $B$ which is due to the contribution ${\cal H}^*_h$. This 
part is diagonal in the vectorial indices and the scalar matrix multiplying the unit 
tensor $\bf I$ reads
\begin{eqnarray} \label{b7}
  B_{mn} =\left( \begin{array}{ccccccc}
      1  & -\nu & 0 & 0 & \cdots & \\*[2mm]
      -\nu   & 1 & -\nu & 0 & \cdots\\*[2mm]
      0 & -\nu   & 1 & -\nu & 0  \\*[2mm]
      \cdots & \cdots & \cdots &\cdots &\cdots \\*[2mm]
\end{array} \right).
\end{eqnarray}
Using the property 
 \begin{equation} \label{b8}
\sin \left[\frac{m (n+1) \pi}{N+1 } \right] + \sin \left[\frac{m (n-1) \pi}{N+1 } \right]=
2  \cos  \left(\frac{m  \pi}{N+1} \right) \sin \left(\frac{m n \pi}{N+1} \right),
\end{equation}
it is easy to see that the eigenvalue equation has the form
\begin{equation} \label{b9}
\sum^N_{n=1} B_{mn} \sin \left(\frac{p n \pi}{N+1 } \right) = 
\theta_p \sin \left(\frac{p n \pi}{N+1} \right),
\end{equation} 
with $\theta_p$ given by
\begin{equation} \label{b10}
\theta_p = \left[1 - 2 \nu \cos  \left(\frac{p  \pi}{N+1}  \right)\right]. 
\end{equation} 
Note that (\ref{b9}-\ref{b10}) are also satisfied for the ends of
the chain ($m,n=0,N$) (which do not correspond to bond indices),
since we can trivially extend the summation to the interval  $0 \leq m,n \leq N$. 

Using Eq. (\ref{b9}) and similar trigonometric relations, we arrive at
the eigenvalues for $M_{mn}$
\begin{equation} \label{b11}
\sum^N_{n=0} M_{pn} \cos \left[\frac{(n+1/2) \pi}{N+1} \right] = \zeta_p 
\cos \left[\frac{(p+1/2) \pi}{N+1} \right],
\end{equation} 
with $\zeta_p$ given by  
\begin{equation} \label{b12}
\zeta _p = 2  \left[1- \cos \left(\frac{p \pi}{N+1} \right)\right] \theta_p.
\end{equation} 

Next we consider the influence of the stabilizing force ${\cal H}^*_s$. Since
\begin{equation} \label{b13}
{\bf B}^s_{i,j} =-  {\partial^2 \over \partial {\bf u}_i \partial {\bf u}_j} \sum^N_{n=1} u_n ,
\end{equation} 
evaluation of the derivative yields
\begin{equation} \label{b14}
{\bf B}^s_{i,j}   =- \delta_{i,j} {1 \over u_i}\left( {\bf I} - {{\bf u}_i {\bf u}_i \over u^2_i }\right).
\end{equation} 
This tensor projects out the longitudinal  eigenmodes and is unity for the transversal
modes. Longitudinal and transversal relate to the direction of the vector ${\bf u}_j$.
Thus the longitudinal modes are given by (\ref{b9}-\ref{b10}). We put this in formula
for the decay constants $\lambda_p$ of the longitudinal mode $p$
\begin{equation} \label{b15} 
\lambda^l_p = \zeta _p. 
\end{equation} 
  
The two transversal modes are corrected by the term (\ref{b14}), which is diagonal
in the matrix indices. The diagonal contribution to $B$ of the quadratic terms in the 
Hamiltonian equals 1. So if we define
\begin{equation} \label{b16}
q_i =(1 - 1/u_i),
\end{equation} 
$q_i$ becomes the diagonal matrix element. The transverse modes for $\bf B$ follow 
from the the diagonalization of this matrix, which has to be carried out numerically. 
One gets an impression of the eigenvalue spectrum by using the bulk value
\begin{equation} \label{b17}
1/u_i \simeq 1 - 2 \nu
\end{equation} 
for each diagonal element. Then the modes of the bond matrix have again the shape
 (\ref{b9}) and the eigenvalues are
\begin{equation} \label{b18}
\theta^t_p = 2 \nu \left[ 1 - \cos \left( { p \pi \over N+1} \right) \right].
\end{equation} 
This formula covers the spectrum closely, provided that we replace for the low modes
$p$ by $p-1$ in the argument of the cosine. Note that upon this substitution the 
eigenvalue for $p=1$ vanishes. This is a general property of the eigenmode in the 
force free case, due to invariance of the energy under an overall rotation.

The dynamical matrix then has transversal eigenmodes similar to (\ref{b12}) and the
eigenvalue equals 
\begin{equation} \label{b19}
\lambda^t _p = 2  \left[1- \cos \left(\frac{p \pi}{N+1} \right)\right] \theta^t_p.
\end{equation} 

An external force changes the shape of the modes and the eigenvalues. The 
transversal mode for $p=1$ no longer vanishes. The effects of a force have to
be incorporated numerically.

\subsection { The persistence length}

The persistence length derives from the orientation correlation function 
\begin{equation} \label{b20}
O_{i,j} = \langle \hat{\bf u}_i \cdot \hat{\bf u}_j \rangle.
\end{equation}
Asymptotically $O_{i,j}$ decays exponentially and the exponent defines the persistence 
length.  It is dominantly dependent on $\nu$ (or $\kappa$). For the large values of 
$\lambda$ (or small values of $T^*$), one gets a good impression of the orientation 
correlation function, by taking the bond lengths as rigid, i.e. by studying the 
extensible jointed chain, for which the orientation correlation function can be 
calculated exactly (see Appendix \ref{bond}). One finds for the persistence length $l_p$
\begin{equation} \label{b21}
{l_p \over a} = -{1 \over \log[\coth(r) - 1/r] } \quad \quad {\rm with} \quad \quad r=b^2 \nu/T^*.
\end{equation}
Note that only $\kappa$ determines this persistence length, since $\lambda$, which
is burried both in $\nu$ and $T^*$, drops out. So a given persistence length sets the ratio 
$\nu/T^*$. This expression gives the persistence length as the number counting the 
monomers over which the orientation persists. 

 \subsection{ The force-extension curve}

The most informative quantity is the force-extension curve. It is the response of the 
end-to-end vector ${\bf L}$ on the external force. It follows from the partition function
\begin{equation} \label{b22} 
Z_f = {\rm Tr} \exp -{\cal H}/(k_b T) 
\end{equation} 
as 
\begin{equation} \label{b23}
\langle {\bf L} \rangle = \langle \sum^N_{n=1} {\bf u}_n \rangle =
{\partial \log Z \over \partial {\bf f}}.
\end{equation} 
This quantity is difficult to simulate, since it fluctuates strongly with a long relaxation
time. The limits of weak and strong forces are however easy to understand. 
For weak forces we  may treat the external force as a perturbation, with the lowest 
order result
\begin{equation} \label{b24}
\log Z \simeq \log Z_0  + {1 \over 2 T^*}  \langle ( [{\bf f} \cdot {\bf L}]^2 \rangle_0.
\end{equation}
The subscript 0 indicates the forceless values. Thus the response is linear in ${\bf f}$: 
\begin{equation} \label{b25}
\langle {\bf L} \rangle \simeq  \langle {\bf L L} \rangle_0 \cdot {\bf f} /T^*= {\bf f} \, 
\langle L^2 \rangle_0 /3 T^*.
\end{equation} 
In the forceless average only the diagonal components survive. 
The mean-squared-average
of the end-to-end vector can be related to the persistence length for chains many times 
longer than the persistence length:
\begin{equation} \label{b26}
\langle L^2 \rangle_0 = 2 N b(0) l_p.  
\end{equation} 
The combination $L_c = N b(0)$ is the contour length  and thus one gets for the initial slope
in the large $N$ limit
\begin{equation} \label{b27}
{\langle {\bf L} \rangle \over L_c } \simeq {2 l_p \over 3 T^*} {\bf f}.
\end{equation} 
One sees from this expression that for low $T^*$ the slope of the rise is steep 
for two reasons: the small factor $T^*$ in the denominator and large $l_p$ in the 
numerator. 
\vspace*{4mm}

\begin{figure}[h]
\includegraphics[width=16cm]{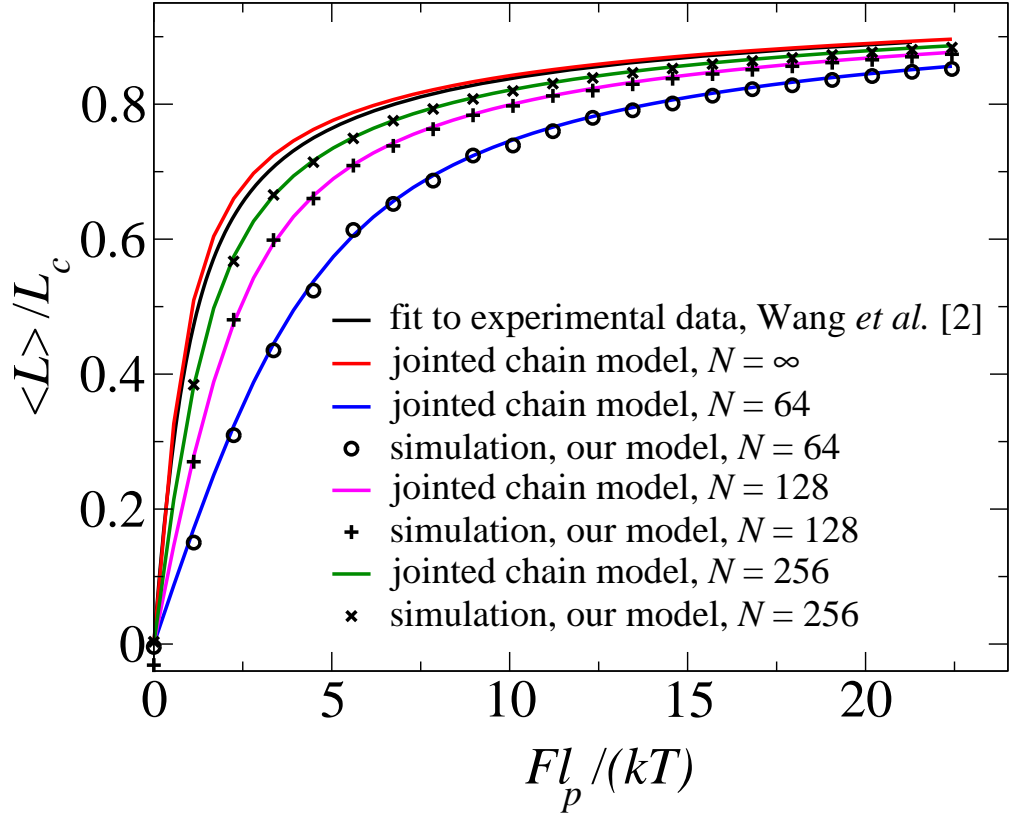}
\caption{Force-extension curves for our model for $N=64,128$ and
$256$, compared to the jointed chain model defined earlier in the
main text for the same $N$-values and to the experimental
data by Wang {\it et al.}  \cite{wang}. The agreement between our
model (simulation) and the jointed chain (calculation) is excellent. 
Note that both models exhibit strong finite-size effects.}
\label{force}
\end{figure}

In the large force limit we may use the result (\ref{b3}). The chain will be stretched in the 
force direction and the bond will be elongated accordingly, with the result
\begin{equation} \label{b28}
 {\langle {\bf L} \rangle \over L_c}  \simeq \, (1+f) \,\hat{\bf f}.
\end{equation} 

As a note in passing we point out on the basis of equation (\ref{b3}) that the ground state 
gives this result for all forces. Thus for $\bf f =0$ there is a residual value for the extension. 
This holds only for $T=0$ as for all $T \neq 0$  the end-to-end vector will vanish for $f=0$,
due to rotational invariance. Indeed the point $(f=0, T=0)$ is singular and the value
of $\langle {\bf L} \rangle$ depends on the way this point is approached.

The complete curve will be a smooth transition between the fast initial rise (till $L$ is
of the order $Nb $) and the slow rise of (\ref{b9}), manifesting itself when $f$
becomes of the order 1. Remember that $f$ is the real force $F$ divided by 
$\lambda d$. Thus forces $f$ of order 1 are huge on a real scale when the polymer
is stiff or $\lambda$ large. In between,  there is a quasi-plateau with the value $Nb$. 
We have shown the force-extension  curve for our model in Fig.~\ref{force} for a number of 
chain lengths $N$.

\subsection{\bf Choice of the parameters}\label{choice}

The model parameters $\nu$ and $T^*$ can be chosen such that the force-extension 
curve of the model is in good agreement with that of biopolymers. Wang et al.~\cite{wang} 
give the following empirical formula for biopolymers, such as dsDNA:
\begin{equation} \label{b29}
 {F l_p \over k_B T} ={1 \over 4} \left[ {1 -  {\langle L \rangle \over L_c} + {F \over K_0}}
\right]^{-2} -  {1 \over 4} +  {\langle L \rangle \over L_c} - {F \over K_0}.
\end{equation} 
This equation contains two polymer-specific parameters: the persistence length $l_p$
and the constant $K_0$. As we have mentioned earlier, the experimental $l_p$ has to be
combined with the bond length $a$ to a dimenionless quantity for which our model 
gives the value (\ref{b21}). For stiff bio-polymers $T^*$ is small, making
$r$ large and $l_p/a$ equal to $r$. So as first equation we get
\begin{equation} \label{b30}
r=\frac{l_p}{a} = \frac{b^2 \nu} {T^*}.
\end{equation}  
The other experimental quantity $K_0$ is turned into a dimensionless combination $z$:
\begin{equation} \label{b31}
z=\frac{a K_0}{k_B T}.
\end{equation} 
This constant manifests itself only for large forces. The behavior of the extension
for large forces is somewhat hidden in the implicit equation (\ref{b29}). In order to 
make it explicit, we write the equation in parametric form. We introduce the auxiliary 
parameter $s$ and for the force parameter $x$:
\begin{equation} \label{b32}
s = {\langle L \rangle \over L_c} - {F \over K_0}, \quad \quad x =  {F \, l_p \over k_B T}
={1 \over 4 (1-s)^2} - {1 \over 4} + s.
\end{equation} 
Thus $x$ is a function of $s$ alone according to equation (\ref{b29}). On the other hand 
the definition of $x$ relates it to $F/K_0$ as
\begin{equation} \label{b33}
x = \frac{F}{K_0} \, rz, \quad \quad {\rm such \quad that } \quad \quad 
{\langle L \rangle \over L_c} = s +  \frac{x}{ r z}.
\end{equation}
The parameter $s$ runs in the interval $0 < s < 1$; small $s$ corresponding to small
forces and $s$ close to 1, to large forces. The last equation (\ref{b33}) gives the
extension as function of the parameter $s$, while the last equation (\ref{b32}) 
gives the force as function of $s$.

For $s=1- \epsilon$, with $\epsilon$ small, one has, according to (\ref{b32}) 
$x \simeq 1/(4 \epsilon^2)$ such that $x \gg s$. 
Then the force-extension curve obtains the form
\begin{equation} \label{b34}
{\langle L \rangle \over L_c} \simeq 1 + \frac{x}{rz}.
\end{equation} 
We compare this with the strong force behavior of our model as given by (\ref{b29}).
Relating the experimental force to our parameter $f$ we have
\begin{equation} \label{b35}
x = {f \, \lambda d \, l_p \over \lambda d^2 \, T^*} = 
{l_p b f \over a T^*} = \frac{r b f}{T^*}.
\end{equation}
The two expressions (\ref{b34}) and (\ref{b28}) are the same if
\begin{equation} \label{b36}
\frac{b}{T^*} = z.
\end{equation} 

We now have two equations, (\ref{b30}) and (\ref{b36}), for the two hamiltonian 
parameters $T^*$ and $\nu$. Solving them leads to the explicit relations
\begin{equation} \label{b37}
T^*=\frac{2r+z}{z^2} \quad \quad {\rm and} \quad \quad \nu=\frac{r}{2 r+z}.
\end{equation} 
For dsDNA the persistence length is about 40 nm. Wang et al.~\cite{wang} report for the 
force constant a value around $K_0=1200$ pN. So with a monomer distance $a=0.33$ nm 
and at room temperature $k_B T = 4 \, {\rm pN \,  nm}$, we get as representative values
\begin{equation} \label{b38}
r=120 \quad \quad {\rm and} \quad \quad z =100.
\end{equation} 
This gives for $\nu$ and $T^*$ as model parameters for dsDNA
\begin{equation} \label{b39}
T^* = 0.034 \quad \quad {\rm and } \quad \quad \nu = 0.35.
\end{equation} 

In Fig.~\ref{force} we show that simulations of our model for large $N$ give a 
force-extension curves in excellent agreement with the curves of the extensible
jointed chain for finite chain-lengths $N=64, 128$ and 256. The extensible 
jointed chain can easily be extrapolated to $ N=\infty$ and this curve follows
closely the empirical curve describing the force-extension of bio-polymers.
So our model is capable to describe bio-polymers that follow the empirical
curve of \cite{wang}. 

 \section{The dynamical Equations} \label{dyn}

We treat the dynamics of the polymer in the so-called large viscosity limit, where
effects of dynamical inertia of the monomers are absent and the solvent is 
represented by a noise source uncorrelated in space and time. So correlations in 
the solvent motion due to hydrodynamical motion are omitted. To incorporate 
these correlations in the noise is another project. Then the dynamical equations 
for the monomers are the  Langevin equations for particles in a high viscous medium:
\begin{equation} \label{c1}
{d {\bf r}_n \over dt} = - {1 \over \xi} {\partial {\cal H} \over \partial {\bf r}_n} 
+{\bf g}_n (t).
\end{equation} 
Here $\xi$ is the friction coefficient and ${\bf g}_n$ is a random noise force,
with correlation function
\begin{equation} \label{c2}
\langle {\bf g}_n (t)\, {\bf g}_m (t') \rangle = {2 k_B T \over \xi} \, {\bf I} \, \delta_{n,m} 
\delta(t-t'),
\end{equation} 
where ${\bf I}  $ is the unit tensor. The strength of the correlation is determined by the 
fluctuation-dissipation relation.

We make the equations dimensionless by introducing a reduced time $\tau$:
\begin{equation} \label{c3}
\tau = t \lambda / \xi.
\end{equation} 
The equations then obtain the form
\begin{equation} \label{c4}
\frac{d {\bf r}_m}{d  \tau} = - \sum^N_{n=0} H_{m,n} \, {\bf r}_n (\tau) +
{\bf h}_m (\tau) +{\bf g}_n (\tau) +{\bf f} (\delta_{n,0} - \delta_{N.n}),
\end{equation}
${\bf h}_m$ derives from the stabilizing term ${\cal H}_s$, given in (\ref{a13}), reading
\begin{equation} \label{c5}
{\bf h}_m = {\partial |{\bf u}|_m \over \partial {\bf r}_m} +{\partial|{\bf u}_{m+1}|  \over 
\partial {\bf r}_m} =|\hat{\bf u}_m| -|\hat{\bf u}_{m+1}|.
\end{equation}
$\hat{\bf u}_n$ is the unit vector in the direction of the bond n. 
The dimensionless random forces have the correlations
\begin{equation} \label{c6}
\langle {\bf g}_n (\tau)\, {\bf g}_m (\tau') \rangle = 2T^* \, {\bf I}  \, \delta_{n,m} 
\delta(\tau-\tau').
\end{equation} 

The center-of-mass position, defined as
\begin{equation} \label{c7}
{\bf R}_{cm} = {1 \over N+1} \sum^N_{n=0} {\bf r}_n,
\end{equation}
is not affected by the interaction between the beads and is only influenced by the 
random forces. It fluctuates due to the random forces as a Brownian particle.

\section{Rouse Modes}\label{rouse}

Diagonalizing the matrix $H_{m,n}$ gives modes relevant for the system. Since 
$H$ is symmetric and positive definite, all its eigenvalues are real and non-negative. 
This means that all the modes, but one, are associated with a decay exponent $\zeta_p$. 
Only the mode corresponding to the center-of-mass has a zero eigenvalue and 
decouples from the dynamic equations (the sum over the columns of $H_{m,n}$ vanishes).

The eigenmodes ${\bf R}_p$ of the matrix $H$  induce an orthogonal transformation
of the positions ${\bf r}_n$. The label $p$ runs through the integers $p=0,1, \dots N$. 
For $p=0$ we have a mode corresponding to the center-of-mass
\begin{equation} \label{d1}
{\bf R}_0 = (N+1)^{1/2} \, {\bf R}_{cm}.
\end{equation}   
The other modes are given by the expression
\begin{equation} \label{d2}
{\bf R}_p = \left({2 \over N+1}\right)^{1/2}  \sum^N_{n=0} 
\cos \left( {p(n+1/2)\pi \over N+1} \right) \,{\bf r}_n.
\end{equation} 
The modes are normalized such that the transformation from ${\bf r}_n$ to ${\bf R}_p$ is 
orthonormal. For that reason we have put the factor in front of ${\bf R}_{cm}$ in (\ref{d1}).
One reckognizes the modes ${\bf R}_p$ as the modes of the Rouse model. The Rouse
modes have generally turned out to be an important tool for analyzing the properties
of a polymer chain \cite{barkema1,barkema2}.
The Rouse modes diagonalize the part ${\cal H}^*_h$ of the hamiltonian.
The eigenvalues are, acoording to (\ref{b13}),
\begin{equation} \label{d3} 
\zeta _p = 2  \left[1- \cos \left({p \pi \over N+1} \right)\right] 
\left[1- 2 \nu \cos \left({p \pi \over N+1} \right) \right],
\end{equation} 
showing that the slow modes vanish as $(p/N)^2$ for small $p$. 

As the Rouse modes diagonalize the dynamic matrix  $H$, it can be written as
\begin{equation} \label{d4} 
\sum^N_{m,n=0} H_{m,n} \, {\bf r}_m \cdot {\bf r}_n = \sum^N_{p=0} \zeta_p \, R^2_p,
\end{equation} 
with $\zeta_p$ given by (\ref{d3}).
We point out  that ${\bf R}_p$ is an exact eigenmode of $H_{m,n}$ because of the special
structure of the boundaries of the the matrix. 

The inverse transformation reads
\begin{equation} \label{d5}
 {\bf r}_n = {\bf R}_{cm} + \left({2 \over N+1}\right)^{1/2}  
\sum^N_{p=1} \cos \left( {(n+1/2)p \pi \over N+1} \right) {\bf R}_p.
\end{equation} 
The advantage of these linear relations between the Rouse modes and the monomer 
positions, is that we can switch between either of two representations of the configuration 
to whichever is more convenient.

By applying the transformation to the equations (\ref{c3}) they become
\begin{equation} \label{d6}
\frac{d {\bf R}_p}{d \tau} = -\zeta_p \,{\bf R}_p (\tau)  +{\bf H}_p + {\bf F}_p + 
{\bf G}_p (\tau).
\end{equation} 
Here ${\bf H}_p$ is the transform of ${\bf h}_n$ defined in (\ref{c5}):
\begin{equation} \label{d7}
{\bf H}_p = \left({2 \over N+1}\right)^{1/2}  \sum^N_{n=0} 
\cos \left( {(n+1/2)p \pi \over N+1} \right) {\bf h}_n.
\end{equation} 
The external force on the Rouse modes gets the form
\begin{equation} \label{d8}
{\bf F}_p = - 2 \cos \left( {p \pi \over 2 (N+1) }\right) \,  {\bf f}.
\end{equation} 

As the transformation to Rouse variables is orthogonal, the correlation function of 
${\bf G}_p$ equals
\begin{equation} \label{d9}
\langle {\bf G}_p (\tau) {\bf G}_q (\tau') \rangle = 2 T^*\, {\bf I} \, \delta_{p,q} 
\delta (\tau-\tau'). 
\end{equation} 
Without ${\bf H}_p$ the Rouse modes are exact eigenmodes with decay coefficient 
${\zeta}_p$. 

We make the scheme more explicit by inserting  (\ref{c5}) into (\ref{d7}) and 
combining the two terms, yielding
\begin{equation} \label{d10}
{\bf H} _p (\tau) =-\sum^N_{n=1} M_{p,n}  \,\hat{\bf u}_n (\tau),
\end{equation} 
where the matrix $M_{p,n}$ is defined as 
\begin{equation} \label{d11}
 M_{p,n} = 2 \left( {2 \over N+1} \right)^{1/2} \sin \left( {p \pi \over 2(N+1)} \right) \, 
\sin \left( {n p \pi \over N+1} \right) .
 \end{equation} 
The bond vectors are likewise expressed in terms of the Rouse variables using (\ref{d5}):
\begin{equation} \label{d12}
{\bf u}_n=-\sum^N_{q=1} {\bf R}_q \, M_{q,n} .
\end{equation} 
By these steps the relation between the stabilizing force and the Rouse modes has been
made operational. We note that the operations have the character of a discrete sine 
transform and therefore the technique of the fast fourier transform can be applied 
in both relations (\ref{d10}) and (\ref{d12}). 

In concluding the discussion of the Rouse modes, we eleborate on the relation between
the modes of the bond matrix $B_{m,n}$, introduced in (\ref{b7}) and the Rouse modes.
To make this relation more transparant, we introduce modified Rouse modes as
\begin{equation} \label{d13}
\tilde{\bf R}_p = - 2 \sin \left( {p \pi \over 2(N+1)} \right) {\bf R}_p.
\end{equation} 
Then relation (\ref{d12}) can be written as
\begin{equation} \label{d14}
{\bf u}_n = \left({2 \over N+1} \right)^{1/2} \sum^N_{p=1} 
\sin \left( {n p \pi \over N+1} \right) \tilde{\bf R}_p.
\end{equation} 
The transformation from the $\tilde{\bf R}_p$ to the ${\bf u}_n$
is also orthogonal and self-dual:
\begin{equation} \label{d15}
\tilde{\bf R}_p = \left({2 \over N+1} \right)^{1/2} \sum^N_{p=1} 
\sin \left( {n p \pi \over N+1} \right) {\bf u}_n.
\end{equation}
This provides actually a short-cut to derive (\ref{d3}). The harmonic energy is on the 
one hand with (\ref{b7}) given by
\begin{equation} \label{d16}
\sum^N_{m,n} B_{m,n} \, {\bf u}_m \cdot {\bf u}_n= \sum^N_{p=1} \theta_p 
\, \tilde{\bf R}_p \cdot \tilde{\bf R}_p
\end{equation} 
and on the other hand it also equals
\begin{equation} \label{d17}
\sum^N_{m,n} H_{m,n} \, {\bf r}_m \cdot {\bf r}_n= \sum^N_{p=1} \zeta_p 
\, {\bf R}_p \cdot {\bf R}_p.
\end{equation} 
With relation (\ref{d13}) between the modified Rouse modes and the Rouse modes,
relation (\ref{d3}) immediately follows.

\section{Rotational diffusion}\label{rotdif}

Apart from the Rouse modes we encountered in section \ref{modes} the mechanical 
modes of the harmonic approximation near the ground state. The ground state is 
relevant for polymers short with respect to the persistence length. Then the majority 
of the configurations will be similar to a ground state. The ground states are 
degenerate with respect to their orientation. A small force lifts this degeneracy.

Both the Rouse modes and
the mechanical modes are linear orthogonal transformations of the monomer
positions and can therefore be expressed in terms of each other. The modes are
similar; in fact the longitudinal mechanical modes are Rouse modes. Also the
Rouse mode $p$ is dominantly present in the representation of the transversal mechanical
mode $p$ in terms of Rouse modes. Nevertheless their role is quite different. 
Each transversal mechanical mode is an exact eigen-mode of the system with
an eigenvalue quite different from the decay constant of the corresponding Rouse 
mode. Since mechanical modes refer to small deviations from the ground state, it
is clear that they do not form an adequate representation, if the ground state is not
dominant (as is the case for polymers longer than the persistence length). 

The orientation of the ground state is best represented by the end-to-end vector
${\bf L}$ defined in (\ref{b23}). It can be expressed in terms of Rouse modes as
\begin{equation} \label{e1}
{\bf L} = -2  \left( {2 \over N+1} \right)^{1/2} 
\sum_{p=1,3, \cdots} \cos \left( \frac{p \pi}{2 (N+1)} \right) {\bf R}_p,
\end{equation} 
showing that the low $p$ modes are most important for the orientation. 
Likewise the mechanical modes with low $p$ will contribute dominantly
in the expression of $\bf L$ in terms of the mechanical modes. 
 
The dynamical equations can also be formulated in terms of the mechanical 
mode-amplitudes $A^a_p (\tau)$. Here the superscript 
$a$ distinguishes the longitudinal ($a=l$) and the transversal ($a=t$) modes. 
As the modes are exact eigen-modes, the equations read
\begin{equation} \label{e2}
\frac{A^a_p (\tau)}{ d \tau} = - \lambda^a_p \, A^a_p (\tau) + G^a_p (\tau),
\end{equation} 
with $\lambda$'s from section \ref{modes}. Note that the equations of the
mechanical modes for different $p$ are independent, since the noise forces 
$G^a_p (\tau)$ are again uncorrelated. Eq.~(\ref{e2}) is a standard Langevin equation for
a variable with a systematic force $\lambda_p$ and a random force $G_p$.
The corresponding Fokker-Planck equation for the distribution is explicitly soluble 
\cite{vankampen}. The decay constant determines the time scale on which the memory of the 
initial value decays.

For longitudinal eigenvalues we have an exact expression, following from 
(\ref{b15}), (\ref{b12}) and (\ref{b10}) as
\begin{equation} \label{e3}
\lambda^l_p =2  \left[1- \cos \left(\frac{p \pi}{N+1} \right)\right] 
\left[1 - 2 \nu \cos  \left(\frac{p  \pi}{N+1}  \right)\right]. 
\end{equation} 
The slow-mode eigenvalues $\lambda^l_p$ decrease with $N$ as
\begin{equation} \label{e4}
\lambda^l_p  \simeq (1- 2 \nu) \frac{p^2 \pi^2}{N^2}.
\end{equation} 
For the transverse eigenvalues we have no closed expression, but they
are well approximated by (\ref{b19}) and (\ref{b18}),
\begin{equation} \label{e5}
\lambda^t_p =4 \nu  \left[1- \cos \left(\frac{p \pi}{N+1} \right)\right] 
\left[1 - \cos  \left(\frac{(p-1)  \pi}{N+1}  \right)\right], 
\end{equation} 
which gives the slow modes the form
\begin{equation} \label{e6}
\lambda^t_p \simeq \nu \, \frac{p^2 (p-1)^2 \pi^4}{N^4}.
\end{equation} 

\begin{figure}[h]
\includegraphics[angle=270,width=\columnwidth]{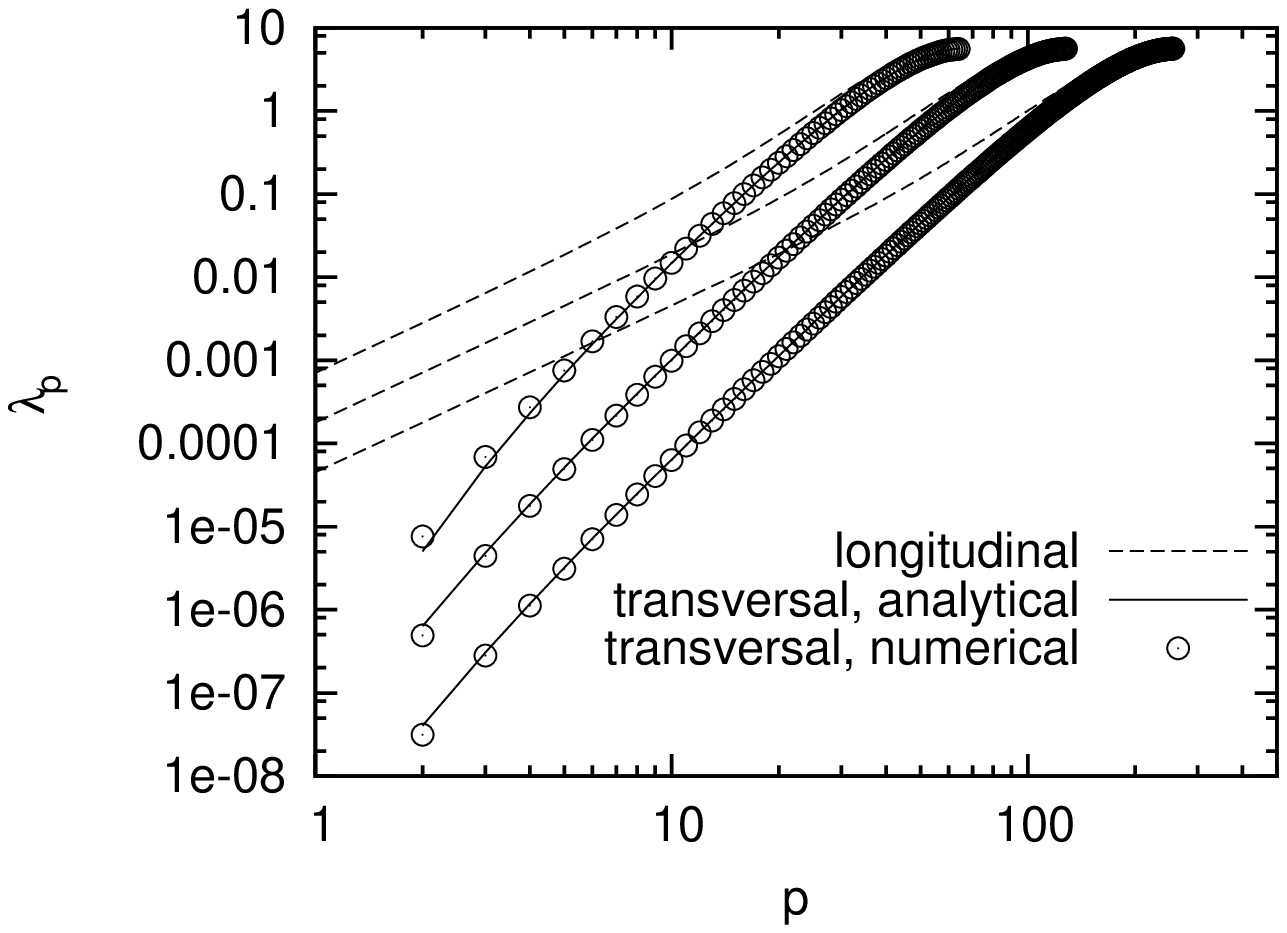}
\caption{Eigenalues $\lambda^l_p$ (dotted lines) and $\lambda_p^t$ (solid lines and points)
as a function of mode number $p$. Here, the model parameters $\nu=0.35$ and $T^*=0.034$ are chosen
to represent dsDNA. The polymer lengths are, from top to bottom, $N=64$, 128 and 256.
The black lines correspond to Eq. (\ref{e5}) for the respective $N$ values.}
\label{decay}
\end{figure}

In Fig.~\ref{decay} we show their behavior as function of mode index $p$ and length $N$.
The striking feature is the large difference in decay coefficient for the slow longitudinal
and transversal modes. The decay as $(N/p)^{-4}$, characteristic for the  
worm-like chain, has often lead to the conclusion that the slowest modes in the system 
show correlations over a time scale increasing as $(N/p)^4$. This is however not the case, a fortiori not for 
transversal mode $p=1$, having a vanishing decay constant.  
If one starts from a ground state with a certain orientation and looks at the evolution on  
a time scale of order $N^3$, then the longitudinal modes have come to equilibrium, 
while the transversal components still fluctuate away from the chosen ground-state 
configuration. In fact they cause a diffusion in orientation, while keeping the shape
of the polymer more or less fixed. The problem of diffusion of a rigid rod has already 
been considered by J.~M.~Burgers in 1938 \cite{rod}. He calculated the rotational
diffusion constant $D_r$ for a rigid rod with moment of inertia $I$ as  
\begin{equation} \label{e7}
D_r = \frac{k_B T}{I \xi} = \frac{12 T^* \lambda}{b^2 N^3 \xi},
\end{equation}
where we have inserted the moment of inertia of the ground state $I=a^2 N^3/12$;
here, we recollect that $a$ is the distance between monomers.
This corresponds with a rotational relaxation time
\begin{equation} \label{e8}
\tau_r = \frac{\lambda}{2 D_r \xi} = \frac{b^2 N^3}{24 T^*},
\end{equation} 
which is indeed of the order $N^3$. Thus on the scale of $\tau_r$, the orientation of the 
original ground state has diffused away and the slow time scale
$1/\lambda^t_p$ of the transversal modes will not be observed. 

It is interesting to compare the decay times $\tau^t_p$ of the transversal modes 
with this rotational-diffusion time scale $\tau_r$. Using (\ref{b21}) we find
\begin{equation} \label{e9}
  \frac{\tau^t_p }{ \tau_r} =\frac{1}{\lambda^t_p \tau_r} \approx 24 \left(\frac{N a}{l_p}\right)
\frac{1}{\pi^4 p^2 (p-1)^2} \ll 1, \quad \quad p >1.
\end{equation}
This formula seems to indicate that $\tau^t_p$ is a factor $N$ larger than $\tau_r$.
However, relation (\ref{e9}) assumes that the persistence length $l_p$ exceeds the 
length $Na$. So the autocorrelation time $\tau^t_p$ for the amplitude of mode $p>1$ is significantly 
{\it smaller} than $\tau_r$. Thus, the scaling of the decay time proportional to $(N/p)^4$
should not be seen as an indication that semiflexible polymers have very long times
for small $p$, but as an indication that their high-$p$ degrees of freedom decay very
quickly.

\section{Discussion}

We have discussed properties of a polymeric chain characterised by a simple 
hamiltonian, in the high viscosity limit with uncorrelated noise. 
We feel that our model fills a gap in the spectrum 
of models with respect to the dynamics of polymers for the following reasons.
\begin{enumerate}
\item With only two adaptable parameters $T^*$ and $\nu$, the model interpolates 
between a worm-like chain and the Rouse model. Depending on the length of the
polymer, short or long with respect to the persistence length, the properties are 
either more worm-like or more Rouse-like. 
This is important for the dynamics of fragments of biological polymers such as dsDNA, 
which neither are worm-like nor are well described by a Rouse model. 
The combination $\nu/T^*$ controls the  persistence length and with 
$T^* \rightarrow 0$ one approaches the worm-like chain.
\item The empirical force-extension curve for biological polymers as proposed by 
Wang et al. \cite{wang} contains also two adaptable parameters: persistence length
$l_p$ and the force constant $K_0$. In Section \ref{choice} we show how to make 
a map between $l_p$ and $K_0$ on the one hand and our model parameters
$T^*$ and $\nu$ on the other hand. The correspondence between the force-extension
curve of our model and that of Wang et al. is excellent, as Fig.~\ref{force} shows.
We point out that our model also enables to discuss the force-extension of 
short polymer fragment as occurring in networks. The appendix contains the details
of such a finite-size analysis. 
\item The equilibrium properties of the model are close to that of the model of Kierfeld
et al. \cite{lipowsky}. The difference is that we use for the bending energy the inner
product of two adjacent bond vectors, while (commonly) the cosine of the angle between the
bonds is used. Our model allows to have the monomer positions as unconstrained dynamical
variables, whereas it is awkward to formulate the dynamics in terms of the angles.
\item The dynamics of the model admits a description in terms of modes, showing 
clearly the wide spectrum of time scales from short for high-indexed modes till very long for
the important low-indexed modes. Two types of modes exist. For polymers much shorter
than the persistence length, the mechanical modes (described in Section \ref{modes})
are relevant. For longer chains the Rouse modes are the more convenient representation.
Since the modes have a wide spectrum of properties we have the opportunity 
to focus on slow and large scale phenomena and to suppress the role of the fast
short-range behavior, which normally greatly slows down the calculations. 
\end{enumerate}

The model has several options for improvement. One could introduce further bond
interactions to make the model more suited to described specific polymers. In the 
same spirit one could add hydrodynamic interactions like in the Zimm model 
\cite{quake, zimm}.
In a subsequent paper we describe further results of simulations of this model for 
realistic parameters $T^*$ and $\nu$  applying to bio-polymers. We there show how
the simulation can be speeded up by orders of magnitude permitting to study
chains of the order of 1000 base pairs, which makes the Rouse modes a convenient
description of the dynamics.  

{\bf Acknowledgement}. The authors are indebted to Debabrata Panja for numerous
stimulating discussions and providing the simulation data of Fig. \ref{force}.

\appendix

\section {The extensible jointed chain} \label{bond}

In this appendix we discuss the properties of the chain in which the bonds all have the
length $b(f)=b(1+f)$. The $f$ dependence of this bond length derives from the 
stretching of the bonds due to the force. Fixing the bond length
reduces the degrees of freedom to the orientations of the bonds, given by
the polar angle  $\theta_n$ between the bond and the direction of the force and the
azimuth angle $\phi_n$ in a plane perpendicular to the force. So the hamiltonian reduces to 
\begin{equation} \label{A1}
{{\cal H} \over k_B T} = {1 \over 2 T^*} \left( N (b(f)-1)^2 -2 \nu  b(f)^2 
\sum^{N-1}_{n=1} \cos \theta_{n,n+1} - 2 b(f) f \sum^N_{n=1} \cos \theta_n \right)
\end{equation}
$\theta_{n,n+1}$ is the angle between the bonds $n$ and $n+1$. 
We observe that, apart from a trivial additive term, the hamiltonian contains two 
dimensionless parameters
\begin{equation} \label{A2}
y  = b(f)^2 \nu /    T^* , \quad \quad {\rm and} \quad \quad w = b(f) f /(2 T^*).
\end{equation} 
The partition function thus is obtained as
\begin{equation} \label{A3}
Z = Tr \exp \left[ y \sum^{N-1}_{n=1} \cos \theta_{n,n+1} + 
2 w \sum^N_{n=1} \cos \theta_n\right],
\end{equation}
where the trace is an integration over all angles $\theta_n,\phi_n$.
For the evaluation we use the expansion in terms of the Legendre polynomials
\begin{equation} \label{A4}
\exp (y z) = \sum_l (2 l +1) \, q_l (y) \, P_l (z),
\end{equation} 
with the $P_l(z)  $ the $l$th Legendre polynomial. The coefficients $q_l (y)$ are recovered
as the integrals 
\begin{equation} \label{A5} 
q_l(y) = {1 \over 2 } \int^1_{-1} dz \exp (y z) \, P_l (z).
\end{equation}  
These elementary integrals, related to the modified half-integer Bessel-functions, can be 
calculated successively using the recurrence relations between the Legendre polynomials.
A Legendre polynomial is subsequently expanded in the bond angles as
\begin{equation} \label{A6} 
(2 l+ 1)\,  P_l (\cos \theta_{n,n+1}) = 4 \pi \sum_m Y_{l,m} (\theta_n, \phi_n) \, 
Y^*_{l,m} (\theta_{n+1}, \phi_{n+1}),
\end{equation} 
with the $Y_{l,m}$ as spherical harmonics. The advantage of this expression is that the 
relative angle between the bonds is expressed in terms of products of function of the
the angles of the bonds in a fixed coordinate system. 

Next we formulate the partition function as a matrix product. 
The first step is the integration over the azimuth angles $\phi_n$. The
angle $\phi_1$ appears only in the first joint between 1 and 2. So its integration 
selects the term $m=0$, as all the higher $m$ integrate to zero. This implies that also the
angle $\phi_2$ disappears from the product, since
\begin{equation} \label{A7}
Y_{l,0} (\theta_1, \phi_1)  Y^*_{l,0} (\theta_2, \phi_2) ={2 l +1 \over 4 \pi} 
P_l (\cos \theta_1) \, P_l (\cos \theta_2).
\end{equation}  
Subsequently one can integrate over $\phi_2$ and so on, each integration over an 
azimuth angle contributing a factor $2 \pi$. 

The remaining integration over the $\theta_n$ is of the type with $z=\cos \theta$
\begin{equation} \label{A8}
T_{k,l} (2 w) = {[(2 k +1)(2 l+1)]^{1/2} \over 2} \int^1_{-1} dz P_k (z) \, \exp (2 w z) \, P_l (z).
\end{equation}
The numerical factor in front of the integral is inserted to yield the property
\begin{equation} \label{A9}
T_{k,l} (2 w) = \sum_n T_{k,n} (w) \, T_{n,l} (w),
\end{equation} 
which follows from the completeness of the Legendre polynomials and which is convenient
for symmetrizing the coming matrix product. 

We now define the transfer matrix
\begin{equation} \label{A10}
S_{k,l} (w,y) =\sum_n  T_{k,n} (w) \, q_n (y) \, T_{n,l} (w) , 
\end{equation} 
which allows us to write the partition function as
\begin{equation} \label{A11}
Z = (4 \pi)^N \, \langle \chi  |\,{\cal S}^{N-1}\, | \chi \rangle,
\end{equation} 
with the definition of the initial state $ | \chi \rangle$
\begin{equation} \label{A12}
| \chi_l  \rangle = T_{l,0}.
\end{equation} 
Note that we have used the property (\ref{A9}) to make the definition of $B_{k,l}$ 
symmetric.

By expression (\ref{A11}) the problem of calculating the partition function is reduced
to finding the eigenvalues of the matrix $S_{k,l}$. For very long chains the largest
eigenvalue will dominate. However, since the persistence length is long, this limit will
only be reached when the chain has several thousands of bonds. Therefore we also
investigate the finite chain for which one has to take into account all the eigenvalues.
The eigenvalues and eigenfunctions are obtained from the equations
\begin{equation} \label{A13}
\sum_n S_{k,n} \psi^\alpha_n = \lambda_\alpha \psi^\alpha_k.
\end{equation} 
Then we use the representation
\begin{equation} \label{A14}
S_{k,n} = \sum_\alpha \psi^\alpha_k \lambda_\alpha \psi^\alpha_n.
\end{equation} 
Note that for small $w$, i.e. for weak forces, the matrix $T_{k,l}$ reduces to $\delta_{k,l}$ 
because the definition is just the orthogonality relation between Legendre polynomials. 
Then the eigenvalues of $S_{k,l}$ are given
by the factor $q_l (y)$ and one of their properties is that 
they decrease with increasing $l$. Thus for $w \simeq 1$, i.e. forces $f \simeq T^*$,  it
suffices to restrict the matrix $S$ to a finite square in the upper-left corner. This simplifies
the problem considerably since, both for the $q_l (y)$ and the $T_{k,l} (w)$, simple recursions
relation exist, enabling to construct them from the first few values. Even for larger $w$
one does not need to include an excessive number of states to get an accurate answer. 

The partition function itself is not the main goal of the above  exercise. We now apply  this 
approach to the calculation of the force extension curve and the orientation correlation 
function. For the former we need the sum of the averages of $\cos \theta_j$ (see (\ref{b7})).   
The appearance of $\cos \theta_j$ means the extra matrix $t_{k,l}$ at position $j$ 
in the matrix product, with
\begin{equation} \label{A15}
t_{k,l} = {[(2 k +1)(2 l+1)]^{1/2} \over 2} \int^1_{-1} dz P_k (z) \, z \, P_l (z).
\end{equation} 
This integral equals
\begin{equation} \label{A16}
t_{k,l} = {k \delta_{k-1,l} + (k+1) \delta_{k+1,l}  \over [(2 k +1)(2 l+1)]^{1/2}}.
\end{equation} 
As we plan to take the eigenfunctions $S$ as basis, we transform the matrix $t_{k,l}$
to that basis:
\begin{equation} \label{A17}
t^{\alpha, \beta} = \sum_{k,l} \psi^\alpha_k t_{k,l} \psi^\beta_l.
\end{equation} 
In the same way we transform the $\chi$ vector:
\begin{equation} \label{A18}
\chi^\alpha = \sum_k \psi^\alpha_k T_{k,0}.
\end{equation} 
The partition function is now obtained as
\begin{equation} \label{A19}
Z = (4 \pi)^N \, \sum_\alpha \chi^\alpha \lambda^{N-1}_\alpha \chi^\alpha.
\end{equation}  
In the infinite long chain limit only the largest eigenvalues $\lambda_0$ contributes.

The numerator of the end-to-end vector gives the sum over all positions of the 
$\cos \theta_j$, or matrices $t$ in the product. We may write this as
\begin{equation} \label{A20}
\langle \sum_j \cos \theta_j  \rangle = {1 \over Z} \sum_{\alpha,\beta} \chi^\alpha 
U_{\alpha, \beta} t^{\alpha,\beta} \chi^\beta.
\end{equation} 
The factor $U_{\alpha,\beta}$ accounts for all the ways the $t$ can be put on the bonds.
On the left of the $t$ one will have the state $\alpha$ and on the right $\beta$. So
$U_{\alpha,\beta}$ is given by
\begin{equation} \label{A21}
U_{\alpha, \beta} = \lambda^{N-1}_\beta + \lambda_\alpha \lambda^{N-2}_\beta +
\cdots  + \lambda_\beta  \lambda^{N-2}_\alpha + \lambda^{N-1}_\alpha.
\end{equation} 
Dividing (\ref{A20}) by $N$ will give the extension, i.e.~the ratio between the
end-to-end vector and the contour length. As mentioned in the limit of an infinite chain
we may restrict the summations over $\alpha$ and $\beta$ to $\alpha=\beta=0$.

For the orientation correlation we consider the infinitely long chain, as its 
definition should not be mudded by finite-size effects. We repeat the definition of the 
orientation correlation for the approximation considered in this appendix:
\begin{equation} \label{A22}
O_{i,j} = \langle \hat{{\bf u}}_i \cdot \hat{{\bf u}}_j \rangle =
\langle \cos \theta_i \cos \theta_j +\sin \theta_i \sin \theta_j  \cos (\phi_i -\cos \phi_j)  \rangle. 
\end{equation} 
The first term gives the contribution of the longitudinal component, the second gives the
transversal component (with respect to the force direction). We restrict ourselves to the
first contribution, since we are mostly interested the persistence length of the force-free 
chain for which the eigenvalue problem is trivial.
Without an external force the longitudinal and transversal components are equal.
The appearance of two terms $\cos \theta_i$ and $\cos \theta_j$ gives an insertion of
the matrix $t_{k,l}$ at the bonds $i$ and $j$. Outside the interval $i-j$ the chain is 
in the state $\psi^0$. Inside the interval it may be in any of the eigenstates $\psi^\alpha$
of $S$. Thus we arrive at the expression for the average
\begin{equation} \label{A23}
\langle \cos \theta_i \, \cos \theta_j \rangle  = \sum_{\alpha} \chi^0 
t^{0,\alpha}  (\lambda_\alpha / \lambda_0)^{|i-j|} t^{\alpha,0} \chi^0.
\end{equation} 
For the force-free chain this expression reduces considerably. As we mentioned for $w=0$
the matrix $T_{k,l}$ reduces to $\delta_{k,l}$ and each $l$ is an eigenstate of $S_{k,l}$
with eigenvalue $q_l (y)$.  Therefore the $t$ matrix with upper indices is the same as
the one with lower indices. Moreover the $t$ matrix couples only nearby states. So the
summation over $\alpha$ is restricted to $\alpha=1$. 
The largest  and the next eigenvalue read
\begin{equation} \label{A24}
q_0 (y) = {\sinh y \over y}, \quad \quad \quad q_1(y) = {\cosh y \over y} -{\sinh y \over y^2}.
\end{equation} 
Therefore the expression for the orientation function reduces to 
\begin{equation} \label{A25}
\langle \cos \theta_i \, \cos \theta_j \rangle  = (q_1(y)/q_0(y))^{|i-j|}/3. 
\end{equation} 
One observes that the correlation decays exponentially with the persistence length $l_p$ 
\begin{equation} \label{A26}
l_p = - {1 \over \log(q_1 (y)/q_0 (y))}  = - {1 \over \log(\coth y - 1/y)} .
\end{equation} 
Large $y$ gives the value $l_p \simeq y$. This has been used in the parameter match between
our model and dsDNA.

\end{document}